\documentclass[lineno]{jfm}

\usepackage{amsmath} 
\usepackage{amssymb}
\usepackage{natbib}
\usepackage{hyperref}
\hypersetup{
    colorlinks = true,
    urlcolor   = blue,
    citecolor  = black,
}
\newcommand{\RomanNumeralCaps}[1]
\linenumbers
\usepackage[normalem]{ulem}
\usepackage{mathtools} 
\usepackage{mathptmx}
\usepackage{multirow}
\usepackage[nice]{nicefrac}
\usepackage{xfrac}
\DeclareMathAlphabet{\altmathcal}{OMS}{cmsy}{m}{n} 

\newcommand{\Relambda}{\textrm{Re}_\lambda}

\title{Clustering in laboratory and numerical turbulent swirling flows}

\author{Sof\'ia Angriman\aff{1,2} \corresp{\email{sangriman@df.uba.ar}},
Am\'elie Ferran\aff{3,4},
Florencia Zapata\aff{1,2},
Pablo J. Cobelli\aff{1,2},
Martin Obligado\aff{3}
\and Pablo D. Mininni\aff{1,2}
}
\affiliation{
\aff{1} Universidad de Buenos Aires, Facultad de Ciencias Exactas y Naturales, Departamento de F\'\i sica. Buenos Aires, Argentina
\aff{2} CONICET - Universidad de Buenos Aires, Instituto de F\'isica de Buenos Aires (IFIBA). Buenos Aires, Argentina 
\aff{3} Universit\'e Grenoble Alpes, CNRS, Grenoble-INP, LEGI, F-38000 Grenoble, France 
\aff{4} Department of Mechanical Engineering, University of Washington, Seattle, Washington 98195-2600, USA}

\begin{document}
\maketitle 
\begin{abstract}
We study the three-dimensional clustering of velocity stagnation points, of nulls of the vorticity and of the Lagrangian acceleration, and of inertial particles in turbulent flows at fixed Reynolds numbers, but under different large-scale flow geometries. To this end, we combine direct numerical simulations of homogeneous and isotropic turbulence and of the Taylor-Green flow, with particle tracking velocimetry in a von K\'arm\'an experiment. While flows have different topologies (as nulls cluster differently), particles behave similarly in all cases, indicating that Taylor-scale neutrally buoyant particles cluster as inertial particles.
\end{abstract}


\section{Introduction}

As is often the case in the study of complex systems, turbulence theory has flourished together with the development of laboratory and numerical methods. For homogeneous and isotropic turbulence (HIT), wind tunnel experiments \citep{Taylor_1938, Shen_2002} and numerical simulations in periodic boxes \citep{Kaneda_2003, Buaria_2019} have played central roles, allowing comparisons with the theory and pushing new ideas. Nonetheless, in non-homogeneous, or multi-phase flows, we lack such benchmarks that would allow a much needed comparison between theory, experiments, and simulations. 

In recent years, the von K\'arm\'an swirling flow in the laboratory \citep{Mordant_2002, Mordant_2004, Volk_2008, Poncet_2008, Huck_2017, Angriman_2020}, and the akin Taylor-Green flow in numerical simulations \citep{Green_1937, Mininni_2011, Kreuzahler_2014}, have provided ways to reach large Reynolds numbers in non-isotropic, non-homogeneous, and in many cases, multi-phase flow regimes. Comparisons of experiments and simulations showed that these flows, albeit differing in the forcing mechanism and boundary conditions, share geometrical \citep{Huck_2017, Mininni_2011, Angriman_2020}, topological \citep{Huck_2017, Angriman_2021}, and statistical properties \citep{Angriman_2020} in both Eulerian and Lagrangian frameworks.

A step forward in these comparisons, that would allow a better understating of flow geometries and that has applications in multi-phase flows, is to study the geometrical properties of zeros and accumulation points \citep{Goto_2006, Goto_2008, Monchaux_2010, Fiabane_2013, Obligado_2014, Mora_2021}.
Moreover, the recent study by \cite{Mora_2021} has performed a combined analysis of nulls spatial properties and inertial particles clustering via Vorono\"i tessellations. The stagnation points, sampled by inertial particles, have also been used to characterize 3D geometrical properties of HIT flows using the Rice theorem \citep{Ferran_2022}.
It is known that even in HIT, nulls of the velocity, vorticity, and of the Lagrangian acceleration do not distribute homogeneously in space, forming clusters \citep{Goto_2006, Mora_2021}. In multi-phase flows, certain particles tend to accumulate preferentially in the vicinity of some of these zeros, affecting their mean free path, and ultimately altering relevant physical processes such as droplet formation in clouds, mixing of chemicals, and phase transitions. The effect of mean flows, and of flow anisotropy and inhomogeneity in the clustering of nulls of the flow vector fields and of particles is unclear and requires further investigation.

Furthermore, not all particles cluster, and it is still uncertain what parameters govern this phenomenon. Very small (i.e., of the size of the Kolmogorov scale) heavy particles display clustering \citep{obligado2019}. No clustering has been observed for small (below the Taylor microscale, but larger than the Kolmogorov scale) neutrally buoyant particles in experiments of HIT \citep{Fiabane_2012}, but clustering has been reported for large particles with sizes comparable to the flow integral scale in von K\'arm\'an experiments \citep{Machicoane_2014, Machicoane_2016}, and attributed to global preferential sampling of the particles of flow inhomogeneities. The intermediate regime, with neutrally buoyant particles of the size of the Taylor microscale has not been studied, and it is unclear whether clustering in such regime would take place, and in that case whether it be caused by large-scale flow sampling effects, or by inertial clustering mechanisms associated to the existence of certain topological points in the turbulent flow.

In this work we perform a numerical and experimental study of the clustering properties of different single- and two-phase turbulent swirling  flows. To this aim, we study and compare the topology of different flows at similar Reynolds numbers (HIT, a Taylor Green flow, and a von K\'arm\'an experiment) using three-dimensional (3D) Vorono\"i tessellations of velocity stagnation points (STPS), zeros of the vorticity (WZEROs), and zeros of the Lagrangian acceleration (ZAPS), as well as the clustering of Taylor-scale  neutrally buoyant particles (PART) in the von K\'arm\'an experiment. Finally, we compare the particles' experimental results with numerical simulations using a minimal model of heavy point particles, to study whether the statistical properties of the clusters of Taylor-scale particles resemble known results of clustering mechanisms in inertial particles.

\begin{figure}
\begin{center}
    {\includegraphics[width=.7\textwidth,trim={0 1.cm 0 0.cm},clip]{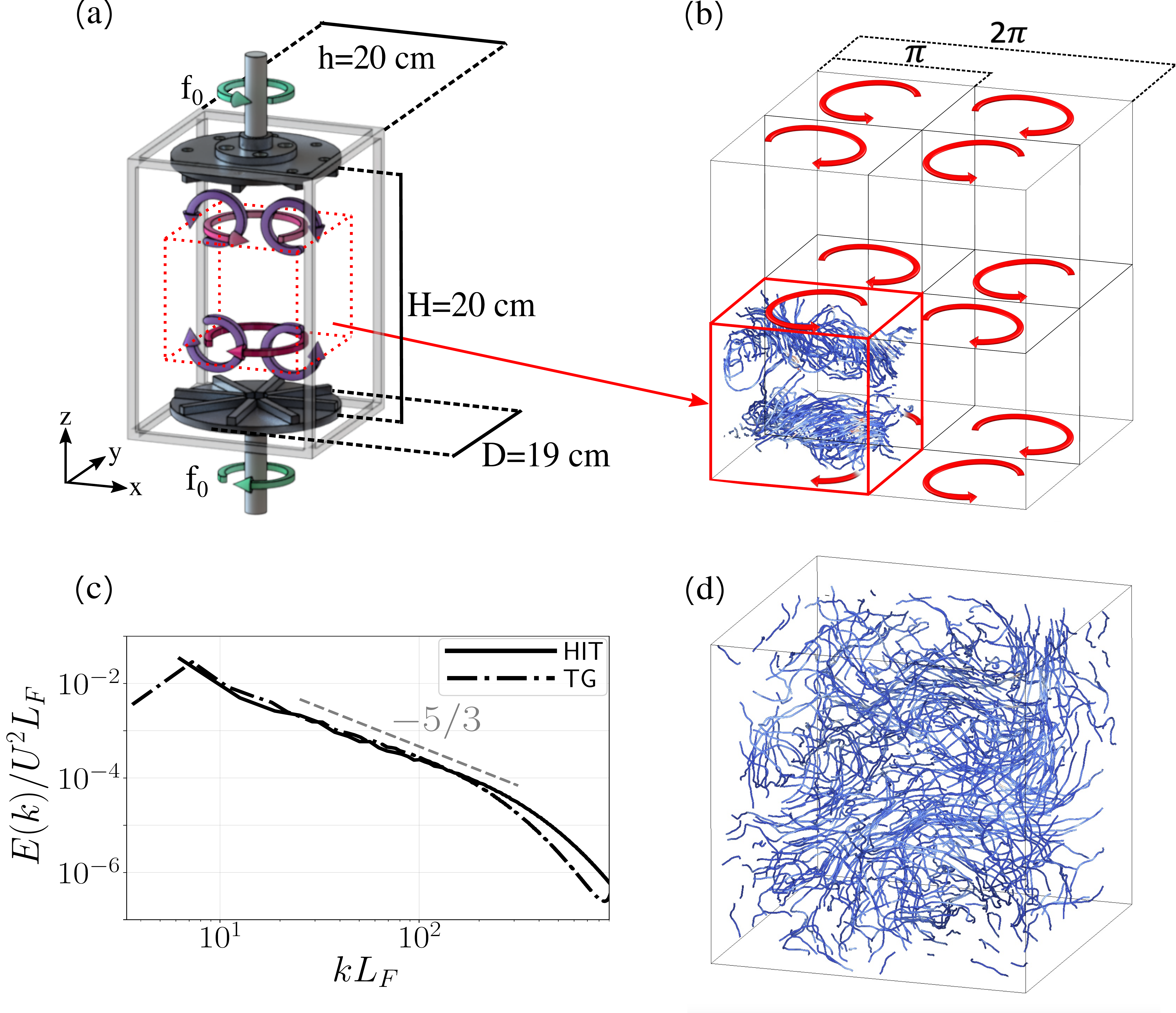}}
\end{center}
	\caption{(a) Experimental set up, and (b) TG and (d) HIT simulations. In the 3D renderings (done with VAPOR; \citealt{Clyne_2007}), the blue lines are streamlines. For the TG simulation in (b), streamlines are shown in one ``VK cell," streamlines in the other cells are similar. The curved red arrows indicate the TG forcing direction. (c) Eulerian energy spectrum for TG and HIT DNSs, Kolmogorov scaling is shown as a reference.}
	\label{fig:diagram}
\end{figure}

\section{Description of the datasets and particle model}

The von K\'arm\'an experiment (denoted herein as EXP) consists of two facing disks of diameter $D=19$~cm (equal to the forcing-based integral scale, $L_\textrm{F}^\textrm{EXP}$), with 8~straight blades each, separated by a vertical distance  of $H=20$~cm. The impellers are in a cell of square cross-section of size $(20 \times 20 \times 50)$~cm$^3$ (room is left at the back of the impellers for shafts to connect them to motors, and for refrigeration coils to remove heat), filled with distilled water from a double-pass reverse osmosis system. The accessible experimental volume, in between the two disks, is thus of $(20\times 20 \times 20)$~cm$^3$. Each impeller is driven by a brushless rotary Yaskawa SMGV-20D3A61 servomotor, with Yaskawa SGDV-8R4D01A servo controllers (see \cite{Angriman_2020} for further details). 
The two impellers rotate in opposite directions with angular velocity $\pm 2\pi f_0$ ($f_0 = 50$ rpm). This generates two large counter-rotating circulation cells producing, on average, a strong shear layer at the mid-plane between the disks. A secondary circulation in the axial direction is also generated by the impellers, resulting in a 3D turbulent flow with an anisotropic large-scale mean flow (depicted schematically in Fig.~\ref{fig:diagram}). We seeded this flow with inertial particles of radius $R = 3$~mm, which are plastic spheres with density $1.02$ times the fluid density, 3D-printed using acrylonitrile butadiene styrene (ABS). The particles radius is such that $R/\lambda = 0.77$ and $R/\eta = 31$, with $\lambda$ the Taylor microscale and $\eta$ the Kolmogorov length scale  (see table \ref{tab:parameters} for definitions). Measurements of particles' dynamics were carried out in a large observation volume $V_{obs}^\text{EXP}$ of size $(16 \times 16 \times 16)~\text{cm}^3$, centered about the geometrical midpoint of the cell, using particle tracking velocimetry (PTV) with two high-speed Photron FASTCAM SA3 cameras at a sampling rate of $125$~fps. 
In total we have $\mathcal{O}(10^5)$ frames. In each frame the mean number of particles simultaneously recorded is $11$, while their maximum recorded number is $29$ (note particles can exit and re-enter the observation volume, exploring also the regions behind the impellers).

\begin{table}
\begin{tabular}{l c c c c c c @{\extracolsep{6pt}} c c c c @{\extracolsep{6pt}} c c c c}
    \multirow{2}{*}{Dataset}  & $\lambda/L_{\text{F}}$ & $\eta/L_{\text{F}}$ & $\Relambda$ & $\textrm{Re}$  & $\textrm{St}$ & $\textrm{Re}_p$ & \multicolumn{4}{c}{$\sigma_{\nu}/\sigma_\text{RPP}$} & \multicolumn{4}{c}{$\langle V_C\rangle ^{1/3}/\eta$}  \\
    \cline{8-11} \cline{12-15}
             & $\times 10^{-2}$ & $\times 10^{-4}$ & & $\times 10^3$ & & & S & Z & W & P & S & Z & W & P\\
    \hline
    EXP & $2.0 $ & $5.0 $ & $430$ & $21$  & $9.95$ & $398$ & - & - & - & 2.33 & - & - & - & 930\\
    TG  & $3.6 $ & $12 $  & $227$ & $6.4$ & $9.41$ & $371$ & 4.09 & 2.69 & 2.81 & 2.84 & 152 & 39 & 22 & 47\\
    HIT & $3.3 $ & $8.7$  & $377$ & $14$  & $9.88$ & 317 & 5.41 & 3.24 & 3.17 & 4.15 & 215 & 49 & 25 & 46
\end{tabular}
\caption{Parameters for the experiment (EXP), and for Taylor-Green (TG) and HIT DNSs. $\lambda = (15 \nu U^2/\varepsilon)^{1/2}$ is the Taylor microscale, with $\nu$ the kinematic viscosity, $\varepsilon$ the energy injection rate, $U$ the r.m.s.~velocity, and $\eta = (\nu^3/\varepsilon)^{1/4}$ is the Kolmogorov length scale, both normalized by the forcing scale $L_{\text{F}}$. $\Relambda = U \lambda/\nu$ and $\textrm{Re} = U L_{\text{F}}/\nu$ are the Taylor and integral scale Reynolds numbers, respectively. $\textrm{St}$ and $\textrm{Re}_p = 2R |\mathbf{u} - \mathbf{v}|/\nu$ are the particles Stokes and Reynolds numbers.
$\sigma_\altmathcal{V}/\sigma_\text{RPP}$ is the ratio of standard deviations of the normalized Vorono\"{i} volumes to those from a random Poisson process, while $\langle V_c \rangle^{1/3}$ is the mean linear cluster size, for STPS (S), ZAPS (Z), WZERO (W) and PART (P).}
\label{tab:parameters}
\end{table}

We also performed direct numerical simulations (DNSs) of the Navier-Stokes equations
\begin{equation}
\partial_t {\bf u} +{\bf u}\cdot{\bf \nabla}{\bf u} = -{\bf \nabla}p +\nu \nabla^{2}{\bf u}+{\bf F},
\end{equation}
where ${\bf u}$ is the solenoidal fluid velocity field ($\nabla \cdot {\bf u}=0)$, $p$ is the pressure per unit mass density, $\nu$ is  the kinematic viscosity, and ${\bf F}$ is an external volumetric mechanical forcing. Equations are written in dimensionless units based on a unit length $L_0$ and a unit velocity $U_0$, and solved in a three-dimensional $2\pi$-periodic cubic box using a parallel pseudo-spectral method with the GHOST code \citep{Mininni_2011, Rosenberg_2020}. A fixed spatial resolution of $N^3 = 768^3$ grid points is used. To mimic the geometry of the large-scale flow in the von K\'arm\'an experiment, the external forcing ${\bf F}$ is given by the Taylor-Green flow,
\begin{equation}
    {\bf F}_\text{TG} = F_0 \left[ \sin(x) \cos(y) \cos(z) \, \hat{x} - \cos(x) \sin(y) \cos(z) \, \hat{y} \right] .
    \label{eq:TG}
\end{equation}
This forcing corresponds to a periodic array of counter-rotating large-scale vortices, which in the domain $[0,\pi)\times [0,\pi)\times [0,\pi)$ reduces to just two counter-rotating vortices separated vertically by a shear layer (see Fig.~\ref{fig:diagram}, where we call this domain a ``VK cell" by analogy with the von K\'arm\'an flow). We also performed a simulation of homogeneous and isotropic turbulence (HIT) with random forcing, to compare against the TG DNS and EXP. The HIT simulation was performed following the same procedures used for the TG DNS, using a spatial resolution of $768^3$ grid points. Turbulence was sustained by injecting energy in all modes in the vicinity of the Fourier shell with wavenumber $ k = 1$, with constant solenoidal amplitude and random phases, i.e., the forcing ${\bf F}$ is given by
\begin{equation}
    {\bf F}_\text{HIT} = F_0 \sum_{|{\bf k}|\in(0,2)} {\mathcal Re} \left\{ \frac{i \: {\bf k} \times \hat{\bf e}_{\bf k}}{|{\bf k}|} e^{i({\bf k} \cdot {\bf r} + \phi_{\bf k})} \right\} ,
\end{equation}
where ${\mathcal Re}$ denotes real part, $\hat{\bf e}_{\bf k}$ is a unit vector, and $\phi_{\bf k}$ is a random phase. The random phases for each mode are slowly evolved in time, with a correlation time of $0.5$ large-scale eddy turnover times. For the TG DNS the magnitude of the forcing wavenumber satisfies $|\mathbf{k}_\textrm{F}^\textrm{TG}| = \sqrt{3}$, so the forcing lengthscale is defined as $L_\textrm{F}^\textrm{TG} = 2\pi/|\mathbf{k}_\textrm{F}^\textrm{TG}| = 2\pi/\sqrt{3}$, while for HIT $|\mathbf{k}_\textrm{F}^\textrm{HIT}| = 1$, and then $L_\textrm{F}^\textrm{HIT} = 2\pi$ (see table \ref{tab:parameters} for relevant parameters).

In both simulations we integrate a minimal model of heavy inertial point particles, satisfying the effective equations of motion
\begin{equation}
    \dot{\bf x}_p = {\bf v}(t) , \qquad
    \dot{\bf v} = [{\bf u}({\bf x}_p,t) - {\bf v}(t)]/{\tau_p},
    \label{eq:iner}
\end{equation}
where ${\bf v}(t)$ and $\tau_p$ are the particles velocity and Stokes time, respectively. 
As the particles radius in EXP is such that the Maxey-Riley approximation is not valid (see, e.g., \citet{Qureshi_2007, Calzavarini_2008, Volk_2008} and \cite{Homann_2010} for detailed studies of limitations of this approximation, and of Eq.~(\ref{eq:iner}) in particular), and as the equations of motion of our particles are not known, Eq.~(\ref{eq:iner}) should be considered as a model with effective parameters, or as a reference model to compare against, to determine whether the clustering of the particles in the experiment resembles the one expected from clustering mechanisms in the case of inertial particles. Note also that, while higher-order terms in the particles radius of the full Maxey-Riley equations could be considered (for instance, considering added mass forces, as done by \citet{Calzavarini_2008}, or keeping Fax\'en corrections \citep{Calzavarini_2009} and Basset-Boussinesq forces \citep{Brennen2005}), the particles in the experiment have $R \gg \eta$, and therefore the perturbative expansion in $R$ in the Maxey-Riley approximation is not valid (indeed, neutrally buoyant particles slightly larger than $\eta$ in such approximation do not cluster, see, e.g., the studies in \cite{Calzavarini_2008} and \cite{Reartes_2021}). We thus regard Eq.~(\ref{eq:iner}) simply as an effective equation with only one tunable parameter. For a brief analysis of how other forces affect the clustering of particles, see Appendix \ref{appendix}. 

The tunable parameter in Eq.~(\ref{eq:iner}) depends on the Stokes number of the particles in EXP. For finite-size particles, it is known that their dynamics can depend on many parameters \citep{Fiabane_2012}, and that the conventional definition of the Stokes number does not properly characterize the particles' dynamics \citep{Xu_2008, Fiabane_2012}. Let us consider the Stokes number $\textrm{St} = \tau_p/\tau_f$, where $\tau_p$ is the particle's response time, and $\tau_f$ is some characteristic time of the flow. Using the standard definition of the viscous relaxation time of the particle \citep{Cartwright_2010}, 
\begin{equation}
  \tau_p^{v} = \frac{2}{9} \left( \frac{\rho_p}{\rho_f} + \frac{1}{2} \right) \frac{R^2}{\nu},
\end{equation}
where $\rho_p/\rho_f$ is the particle-to-fluid mass density ratio, and setting $\tau_f$ as the Kolmogorov time-scale $\tau_\eta$, for the particles in EXP we obtain $\textrm{St}_v^{\text{EXP}} = \tau_p^v/\tau_\eta = 312$. This choice would result in particles almost decoupled from the fluid (i.e., with too much inertia) if used in Eq.~(\ref{eq:iner}).

To calculate an effective $\tau_p$ for EXP we then set the Stokes number as $\textrm{St} = \tau_p/\tau_\eta$ and we estimate an effective particle time $\tau_p^\textrm{EXP} = (R^2/\varepsilon)^{1/3}$ as the turbulent turn-over time at the particle radius \citep{Angriman_2020}. 
This choice can be understood as follows. For a spherical particle of radius $R$ and mass $m_p$ moving with velocity $\mathbf{v}$ in a uniform velocity field $\mathbf{u}$, the drag force $\mathbf{F}_D$ acting on the particle per unit mass can be estimated as \citep{Batchelor_2000}
\begin{equation}
    \frac{\mathbf{F}_D}{m_p} = \frac{1}{2}~ \frac{\rho_p}{m_p}~ S~ C_D(\text{Re}_p)~|\mathbf{u} - \mathbf{v}|~(\mathbf{u} - \mathbf{v}),
    \label{eq:drag_force}
\end{equation}
where $\rho_p = 3 m_p/(4\pi R^3)$ is the particle's density and $S = 4\pi R^2$ represents its surface. $C_D(\text{Re}_p)$ is the drag coefficient, which is a function of the particle Reynolds number, defined as $\text{Re}_p = 2 R |\mathbf{u} - \mathbf{v}|/\nu$.
Then, Eq.~(\ref{eq:drag_force}) reads
\begin{equation}
  \frac{\mathbf{F}_D}{m_p} = \frac{3}{2}~C_D(\text{Re}_p)~ \frac{|\mathbf{u} - \mathbf{v}|}{R}~(\mathbf{u} - \mathbf{v}), 
\end{equation}
which can be approximated as $\mathbf{F}_D/m_p \approx (\mathbf{u} - \mathbf{v})/\tau_p$, with
\begin{equation}
    \tau_p \sim \frac{2}{3}~\frac{1}{C_D(\text{Re}_p)}~\frac{R}{\langle |\mathbf{u} - \mathbf{v}| \rangle},
    \label{eq:taup_0}
\end{equation}
where angle brackets denote time averages.
For finite-size neutrally buoyant particles,
\citet{Cisse_2013} estimated $\langle |\mathbf{u} - \mathbf{v}| \rangle \sim (\varepsilon R)^{1/3}$,  assuming Kolmogorov scaling and proposing a self-similar solution in the surroundings of the particle, and verified the validity of this estimation using DNSs. Substituting their result in Eq.~(\ref{eq:taup_0}) leads to
\begin{equation}
    \tau_p \sim \frac{2}{3}\frac{1}{C_D(\text{Re}_p)} ~ \left(\frac{R^2}{\varepsilon}\right)^{1/3}.
    \label{eq:taup_1}
\end{equation}
In EXP we can also estimate the particle-fluid slip velocity $|\mathbf{u} - \mathbf{v}| \approx (|\langle u \rangle^2 - \langle v \rangle^2|)^{1/2}$ from the 3D r.m.s.~flow and particles velocities \citep{Bellani_2012}, which gives $\text{Re}_p \approx 400$. For this value of $\text{Re}_p$, $C_D \approx 0.6$  \citep{Batchelor_2000, Morrison_2013}, and then the dimensionless pre-factor $2/[3 C_D(\text{Re}_p)] \approx 1.1$; thus, the particle response time $\tau_p^{\text{EXP}} = (R^2/\varepsilon)^{1/3}$ can be interpreted as an effective time that takes into account the mean effect of  non-linear drag corrections. This in turn yields $\textrm{St}^\text{EXP}=9.95$. Other alternative definitions of the Stokes number have been introduced in the literature. For instance, \citet{Xu_2008} and \citet{Schmitt_2008} propose to use $\tau_p = \tau_p^{v*} = (1/18) (\rho_p/\rho_f) (4R^2/\nu)$, and to set $\tau_f$ as the turbulent dynamic time at the scale of the particle, i.e., $\tau_f = (4R^2/\varepsilon)^{1/3}$.  For the EXP data this yields $\textrm{St}^\text{EXP}_{v*} = \tau_p^{v*}/\tau_R = 14$, which is of the same order of magnitude as our estimation.

In the DNSs, $\tau_p$ is set equal to $\tau_p^{EXP}$ in units of $\tau_\eta$ (i.e., we set the same Stokes number $\textrm{St}$ within statistical fluctuations). In other words, given the values of $\tau^\text{DNS}_\eta$ in the simulations, and of the Stokes number $\textrm{St}^\text{EXP}$ in the experiment, we set $\tau_p$ in Eq.~(\ref{eq:iner}) as $\tau^\text{DNS}_p = \text{St}^\text{EXP} \tau^\text{DNS}_\eta $. An effective radius of the particles in the DNSs can then be estimated consistently with Eq.~(\ref{eq:taup_1}) as $R = [\varepsilon (\tau_p^\textrm{DNS})^3]^{1/2}$, and to compute the value of $\text{Re}_p$ reported in table \ref{tab:parameters} for the simulations we use $\mathbf{u}$ and $\mathbf{v}$ for each particle at every instant. Note both the experimental and numerical estimations of the Reynolds particle number yield Re$_p \gg 1$, reinforcing the fact that models for particles of moderate size do not apply in our case, and that the equations for inertial particles in the DNSs should be understood only as a crude effective model. All the values of St and Re$_p$ for EXP, TG and HIT data are listed in table~\ref{tab:parameters}.

\section{Results}

\subsection{Vorono\"{i} volumes probability distribution}

Nulls of all vector fields were computed in the DNSs using the methods described in \citet{Haynes2007} and \citet{Mora_2021}. Using the positions of the zeros, and the particles' positions in the experiment and DNSs, 3D Vorono\"i tesselations were computed for all datasets. While the DNSs use $10^6$ particles, in the experiment we have a maximum of $\approx 29$ particles in each frame, resulting in a significantly smaller number of closed 3D Vorono\"i cells: there are an average of 3 closed cells per snapshot. When considering all snapshots this results in a total number of closed cells of $1.9\times10^5$. Also, in DNSs, the tessellations were performed with periodic boundary conditions. These two observations will be relevant to interpret some of the results.

\begin{figure}
    {\includegraphics[width=1\textwidth]{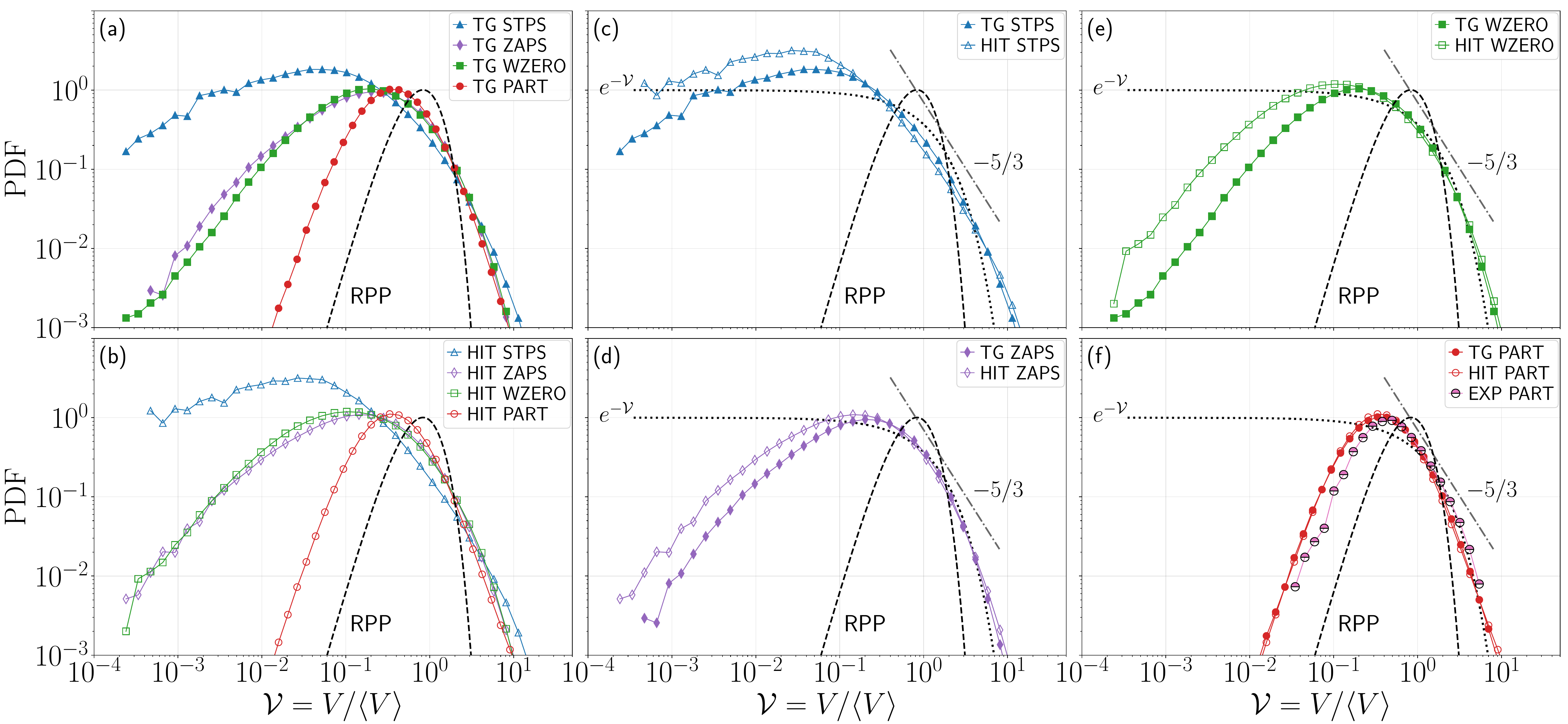}}
	\caption{Probability density functions (PDFs) of normalized Vorono\"{i} volumes, $\altmathcal{V} = V/\langle V \rangle$, for STPS, ZAPS, WZERO, and inertial particles positions (PART). Panels (a) and (b) show a comparison of all quantities in TG and HIT. PDFs of (c) STPS, (d) ZAPS, and (e) WZERO for TG and HIT. (f) PDFs of PART for TG, HIT, and EXP. A Random Poisson Process (RPP), an exponential, and a $-5/3$ power law are shown as references.}
  \label{fig:vor_volumes}
\end{figure}

Figure \ref{fig:vor_volumes} shows the PDFs of the normalized volumes of the Vorono\"i cells, $\altmathcal{V} = V/\langle V \rangle$, for STPS, ZAPS, and ZWEROs in TG and HIT, and the PDFs of the Vorono\"i cells for the particles in the DNSs and the experiment.  For the experimental data, since the total number of particles fluctuates in time (as particles can enter or exit the observation volume $V_{obs}^\text{EXP}$, and as already mentioned, can explore regions behind the disks), $\langle V \rangle_\text{EXP}$ is defined as the volume occupied by the fluid between the propellers divided by the mean number of particles per image, calculated from all of the available images. This choice is robust to variations in the number of particles \citep{Fiabane_2013}, and to the presence of spurious volumes near the boundaries of the images as will be discussed later.

It has been hypothesized that the forcing mechanism or the large-scale flow can introduce changes in the clustering of nulls of the fluid vector fields \citep{Goto_2009}. This figure confirms this is indeed the case: HIT displays stronger clustering of all fields than TG (i.e., larger probabilities for small volumes). In the figures, the PDF resulting from a random Poisson process (herein RPP, corresponding to homogeneous distribution of nulls or particles) is shown as a reference. The larger standard deviation of the PDFs (compared to the RPP, whose standard deviation is $\sigma_\text{RPP} \approx 0.42$) and the stronger tails (of clusters to the left, and of voids to the right) indicate enhanced clustering. However, the Vorono\"i cell volumes of the particles is similar for the three flows, and notably, the level of clustering as quantified by $\sigma_\altmathcal{V}$ is similar for TG and EXP (see table~\ref{tab:parameters}). As will be discussed next, differences between EXP and TG can be associated with finite sampling and with boundary effects.

In Fig.~\ref{fig:vor_volumes} we also show an exponential and a -5/3 power law to serve only as references for the comparison. Note STPS in HIT are more similar to the $\altmathcal{V}^{-5/3}$ power law in the vicinity of $\altmathcal{V} \approx 1$ (i.e., for intermediate volumes, their distribution is closer to a power law than to an exponential decay), which is a consequence of the power-law behavior of the velocity auto-correlation function \citep{Davila_2003, Smith_2008}. TG STPS display a less clear scaling, which is compatible with contamination of the scaling of the velocity by the mean flow \citep{Angriman_2020}. ZAPS and WZERO do not exhibit a power law behavior, and are closer to an almost exponential decay $\sim e^{-\altmathcal{V}}$ in all DNSsfor large volumes (i.e., for $\altmathcal{V}>1$), with the aforementioned differences between HIT and TG for very small values of $\altmathcal{V}$. Remarkably, the PDFs of Vorono\"i cells of all particles are very similar in spite of the differences in the distribution of zeros, and for intermediate and large volumes ($\altmathcal{V} \gtrsim 1$) are closer to an exponential decay, indicating the clustering of particles for those volumes is more akin to the distribution of either ZAPS or WZERO than to the distribution of STPS. It is worth recalling that some studies of finite-size, neutrally buoyant particles in HIT laboratory flows indicate that small particles do not cluster and behave as tracers \citep{Fiabane_2012}, at least from a Vorono\"{i} tessellation point of view. \citet{Fiabane_2012} considered neutrally buoyant particles whose radii ranged from $2.25\eta$ to $8.5\eta$ ($0.0875\lambda$ to $0.23\lambda$). The particles employed here are well outside this range, with a size comparable to the Taylor microscale, a scale at which viscous effects from the turbulent flow become less relevant. Our results suggest that Taylor-scale neutrally buoyant particles cluster similarly to inertial particles. The numerical particle data thus serve the purpose of gauging the level of clustering the experimental particles experience when compared to the minimal model in the simulations. As particles in the DNS cluster akin to different flow nulls, the comparison allows to further relate the clustering in the experiment with specific properties of the background flow.

\subsection{Biases in the evaluation of Vorono\"{i} tessellation using particles' positions}\label{sec:biases}

Measurement of 3D clustering in experiments poses multiple challenges. Particle seeding must be sufficiently small to remain in the regime of a dilute suspension \citep{Elghobashi_1994}, which results in a few particles per measurement. When computing 3D Vorono\"{i} tessellations, this results in a small number of closed cells, roughly proportional to the cube root of the total number of detected particles. Thus, the number of closed Vorono\"{i} cells scales slowly with the number of particles (see \citealt{Tagawa_2012} for another study of 3D clustering in experiments). Lastly, boundaries have a drastic effect in the statistics. In this section we compare DNSs, synthetically post-processed DNSs, and laboratory data, to quantify the effect of these ever-present limitations. We therefore evaluate possible sources of biases in the 3D Vorono\"{i} tessellation calculated from the inertial particles' instantaneous positions by processing DNS data so as to mimic experimental conditions.

\begin{figure}
    \begin{center}
    {\includegraphics[width=0.49\textwidth]{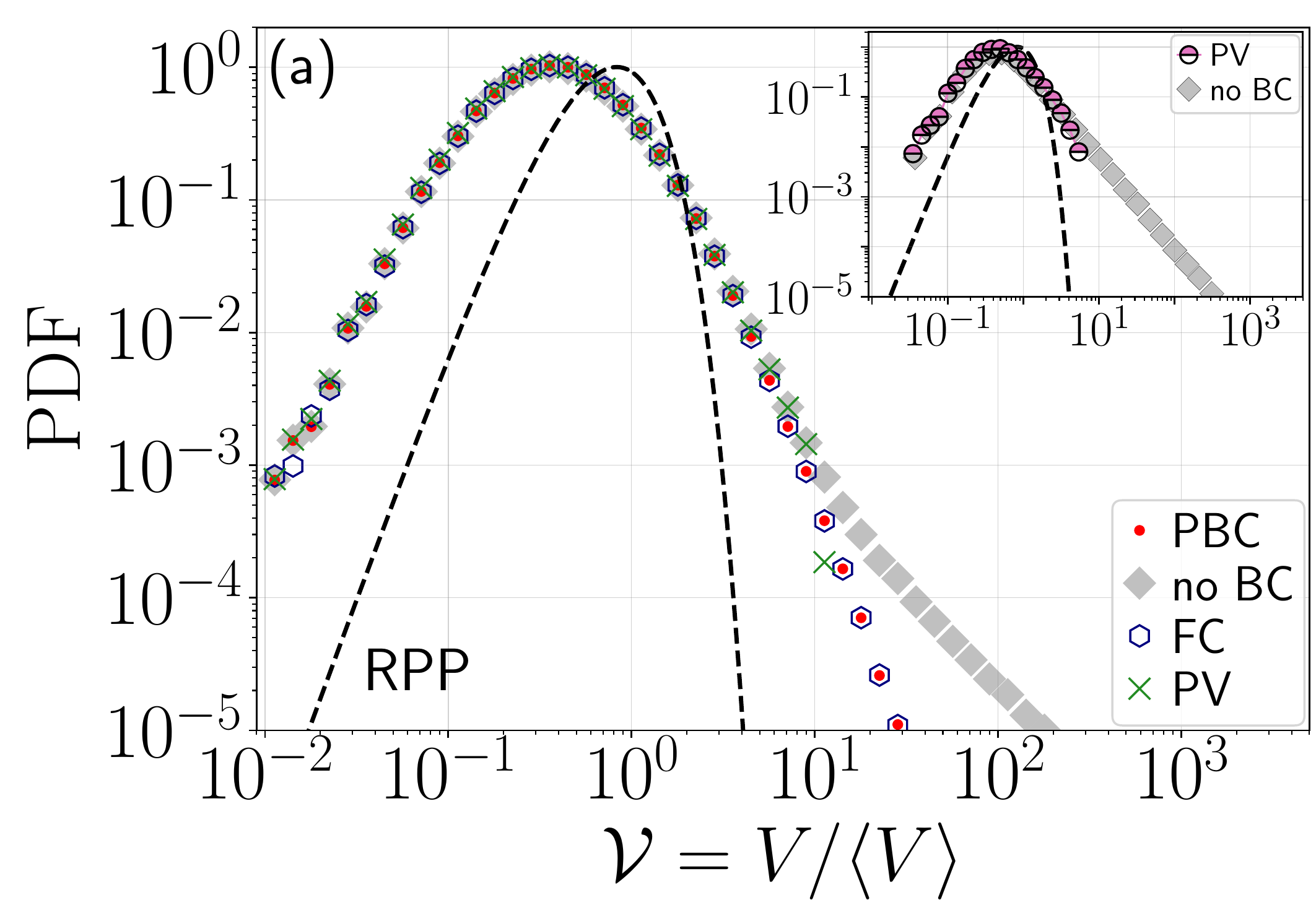}}
    {\includegraphics[width=0.49\textwidth]{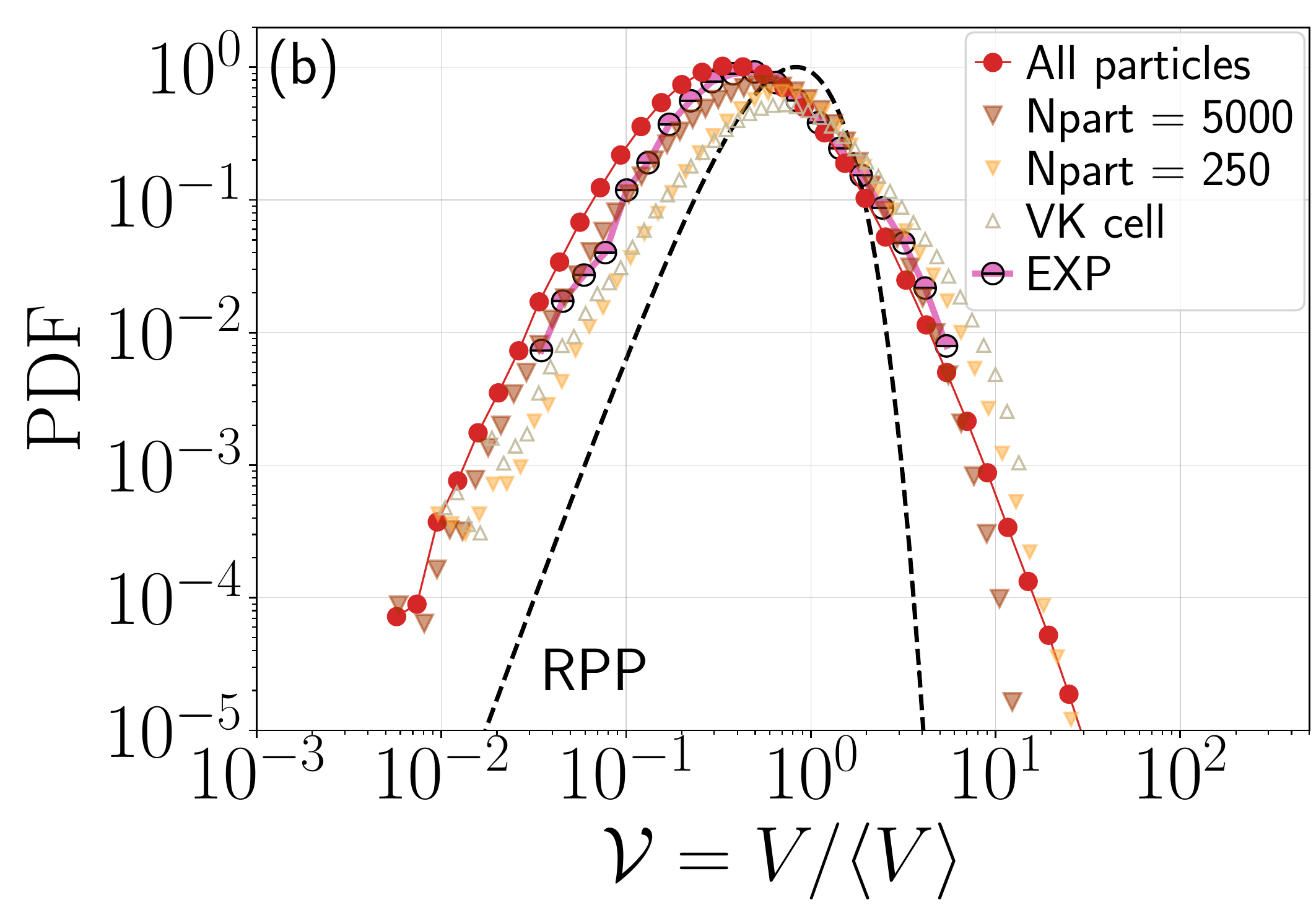}}
    \end{center}
	\caption{
    (a) TG data tessellated without any boundary conditions (no BC), with periodic boundary conditions (PBC), removing frontier cells (FC), or keeping only physical volumes (PV). Inset shows EXP data. Note the spurious tails for large $\altmathcal{V}$.
	(b) TG data sub-sampled to 250 or 5000 particles, and sub-sampled in only a VK cell with an average of 13 particles, is compared to the tessellation keeping all particles and to EXP data. Note the lower probability of small volumes when sub-sampling.
	}
  \label{fig:bias}
\end{figure}

We first focus on the effect of boundaries. Particles near the boundaries of the observation region in the experiment occupy either open cells (which are always discarded in the analyses) or artificially large cells, as particles closing their cells (or the domain walls) are not visible in the cameras field of view. The result is depicted in Fig.~\ref{fig:bias}(a) (inset), which shows the Vorono\"i tessellation on the raw EXP data without imposing any restriction on the cells near boundaries (labeled ``no BC" in the figure); note the power-law tail for very large $\altmathcal{V}$. We reproduced this situation in the TG DNS by computing the tessellation imposing periodic boundary conditions (``PBC", considered as ground truth), by imposing no condition on the boundaries (``no BC", which results in a power-law dependence), by removing all cells close to boundaries (labeled ``FC" in the figure), and by dropping the largest cells that resulted in a total volume larger than the observation volume (that is, keeping only ``physical" volumes, ``PV"), see Fig.~\ref{fig:bias}(a). The latter method, when applied to EXP data, results in the PDF shown in the inset of Fig.~\ref{fig:bias}(a) (the strategy also used in Fig.~\ref{fig:vor_volumes}(f)). 
In each case, $\langle V \rangle$ is the total volume occupied by the fluid divided by the mean number of particles, to remove biases (especially when large spurious volumes are kept). Boundaries mainly affect the statistics of large cells, but the left tails of the PDFs remain the same regardless of the strategy used for treating the boundaries. This suggests boundaries do not affect significantly the determination of clusters (as clusters are formed by small volumes $\altmathcal{V}$ below some threshold, as will be discussed in the following section). Furthermore, we verified that when analysing DNS data in the same way as EXP data (i.e., by considering the non-periodic VK cells in the TG DNS and keeping only ``physical" Vorono\"i cells), the distribution of cluster volumes reported in the following section remains unchanged.

The relatively low number of particles in experiments also has an effect in the PDFs. Figure \ref{fig:bias}(b) shows EXP data together with the original TG data (tessellated using periodic boundary conditions), and the TG data sub-sampled to a maximum of 250 and 5000 particles. 
These sub-sampled sets were built so as to follow the same distribution of number of particles per frame as in the experiments. This was done in the following way. The maximum number of particles $N_\text{max}$ considered (either 250 or 5000) represents the DNS-equivalent of the total number of particles inside the experimental vessel (but not necessarily within the observation volume at once); e.g. $N_\text{max} = 250$ corresponds to roughly 31 particles available in each of the 8 individual VK cells the DNS comprises. The experimental variability in the actual number of particles that simultaneously occupy the observation volume at any time instant was mimicked in these numerical datasets based on the occupation probability distribution (i.e., the probability to observe a certain number of particles at each time instant), which was calculated empirically from the EXP data. As a result, for $N_\text{max} = 250$, the mean number of particles per frame in the complete numerical domain of volume $(2\pi)^3$ is $\approx$ 111, leading to approximately 13 particles per VK cell of the TG flow, comparable to what is attained in the experiments. These sub-sampled datasets were then processed in a manner identical to what was done with the experimental data; namely, employing a tessellation scheme without considering periodic boundary conditions and keeping only physical volumes in the entire $(2\pi)^3$ domain. To further mimic the experimental conditions, the sub-sampling of the TG data was also performed in one VK cell, i.e., in just the octant comprised by the subdomain $[0,\pi)\times[0,\pi)\times[0,\pi)$, thus keeping an average of only $13$ particles per snapshot. The tessellation was then carried out as in the case of $N_\text{max} = 250$ and $N_\text{max} = 5000$ but in the smaller subdomain; the resulting PDF is shown in Fig.~\ref{fig:bias}(b) as well.

Decreasing the number of particles in the DNS in these ways -in both the ($2\pi)^3$ domain and in the VK cell- as shown in Fig.~\ref{fig:bias}(b), results in PDFs whose left tails slowly move towards the RPP, partially explaining the slight defect of small volumes $\altmathcal{V}$ seen in the EXP data. The reduction in the number of particles, and the treatment of boundaries discussed above, also has a small effect in the concavity of the right tail of the observed PDF. For a detailed study of other experimental biases in 2D Vorono\"i tessellation, including sub-sampling effects, see \citet{Monchaux_2012}, and for an investigation of finite-size effects on Vorono\"i analysis of randomly placed spheres, see \citet{Uhlmann_2020}. 

\subsection{Effect of preferential sampling of large-scale inhomogeneities in the experiment}\label{sec:pref}

\begin{figure}
    \begin{center}
    {\includegraphics[width=0.65\textwidth]{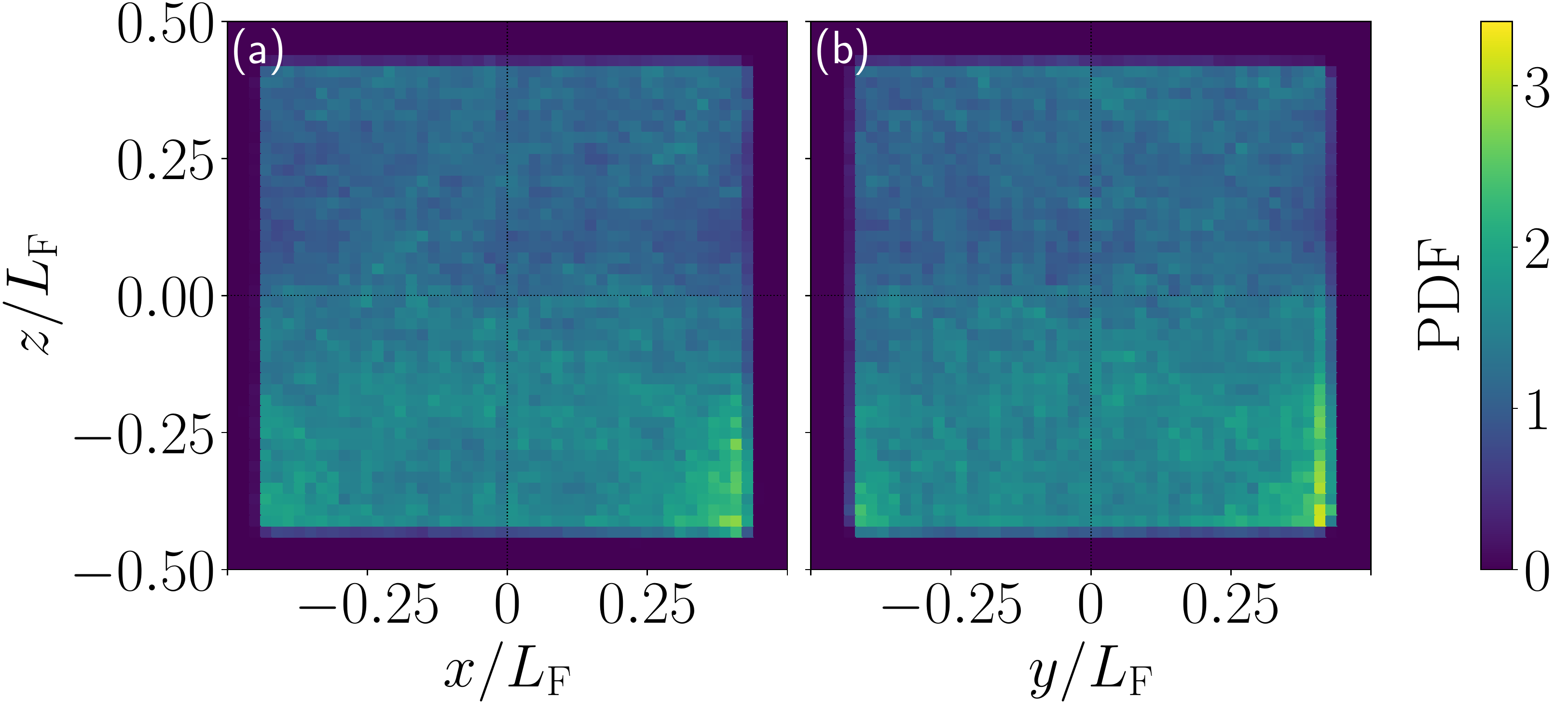}}
    {\includegraphics[width=0.33\textwidth]{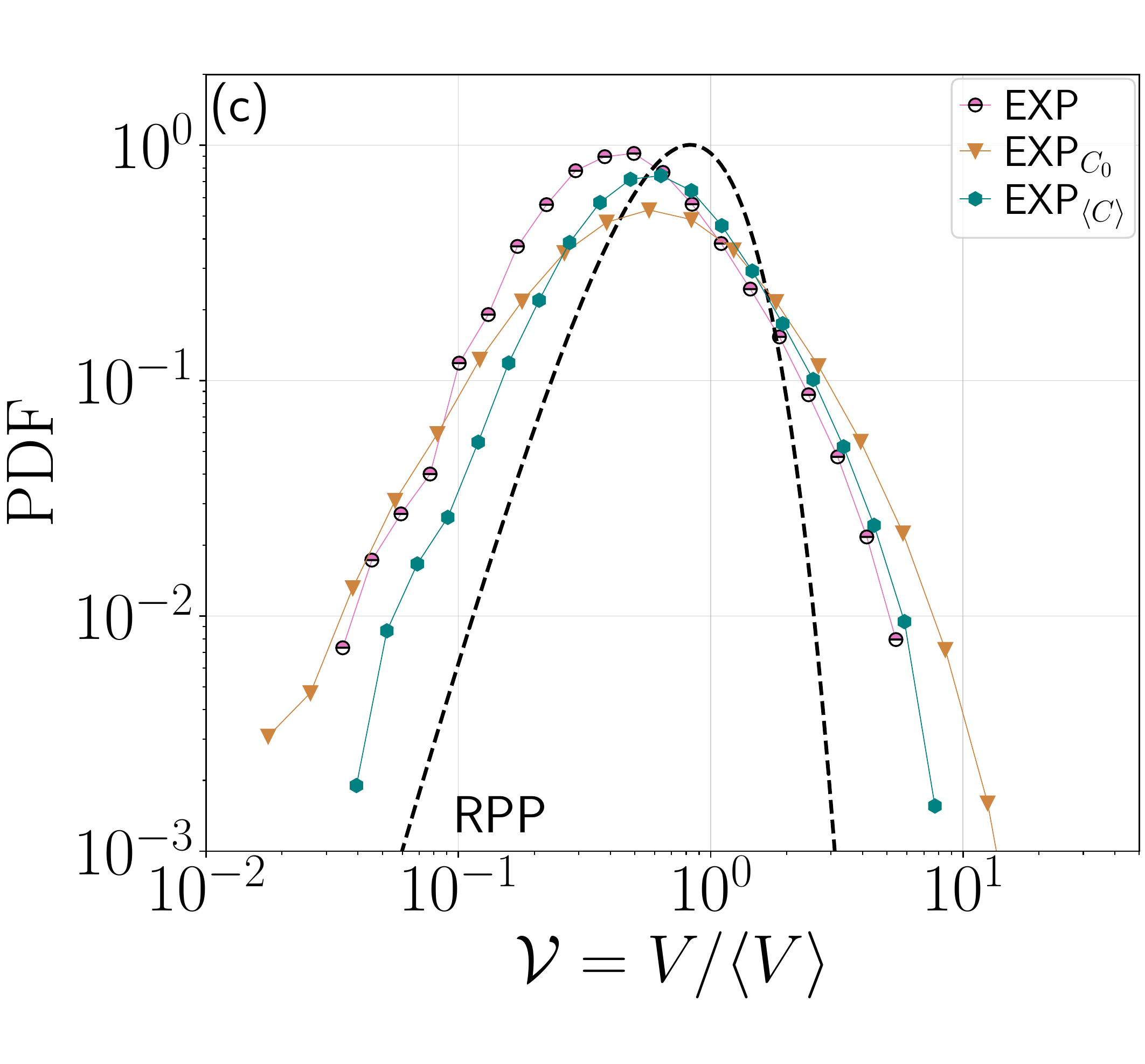}}
    \end{center}
	\caption{Two-dimensional PDFs of inertial particles positions in EXP, (a) $x-z$ plane, and (b) $y-z$ plane, where $\hat{z}$ is parallel to the axis of rotation of the propellers. (c) PDF of the original normalized Vorono\"i volumes for EXP, same PDF compensating each volume by the concentration at the corresponding particle position $C_0=C({\bf x}_0)$ (i.e., PDF of $\altmathcal{V}C_0$), and compensating by the average concentration in each volume $\langle C \rangle=\langle C ({\bf x})\rangle_\altmathcal{V}$.}
  \label{fig:cmap_VK}
\end{figure}

While small (larger than the Kolmogorov scale) particles do not cluster \citep{Fiabane_2012}, neutrally buoyant particles with sizes of the order of the flow integral scale have been reported to cluster in von K\'arm\'an experiments \citep{Machicoane_2014}. The clusters in that case were found to be associated to large-scale flow inhomogeneities (i.e., preferential sampling by the particles) and to confinement effects \citep{Machicoane_2016}, rather than to inertial effects. It is then worth analyzing if the clustering behavior observed in our finite-size particles is the result of, or at least affected by, a global preferential sampling as the one reported by \citet{Machicoane_2014}. To this end we computed the 3D spatial concentration map $C({\bf{x}})$ of the particles in the experiment, using all available snapshots and normalized so that the average in the entire observation volume is $\langle C({\bf{x}})\rangle_{V_{obs}^\text{EXP}} = 1$. Figure \ref{fig:cmap_VK}(a) and (b) show two-dimensional probability density maps of the particles' positions (proportional to the 2D concentration maps) in the $x-z$ and $y-z$ planes respectively, where $\hat{z}$ is chosen parallel to the rotation axis of the propellers and to the direction of gravity. Our particles present a more uniform distribution than the larger particles considered in \citet{Machicoane_2014} and \citet{Machicoane_2016}. This is consistent with previous observations: in \citet{Machicoane_2014} the larger particles are the ones that display the stronger inhomogeneous preferential sampling, while their smaller particles (with $R /\lambda \approx 2.6$) sample the flow more uniformly than their larger particles. In spite of this, in Fig.~\ref{fig:cmap_VK} a very small tendency for our particles to be in the lower half of the cell ($z < 0$) can still be observed, specially near the corners. This may be partially due to asymmetries in the position of the shear layer in the von K\'arm\'an flow \citep{Ravelet_2004, Huck_2017}, to a small effect of gravity, and to the particles entering and exiting the region behind the propellers; the latter effect is particularly visible in the lower corners of the concentration maps.

To quantify the effect that the particles preferential sampling has on the observed clustering, we computed the Vorono\"i cells compensating each cell volume $\altmathcal{V}$ by its local particle concentration. 
The purpose of doing this is to correct any underestimation (resp. overestimation) of the cell volumes resulting from the cells being located in regions of space with higher (resp. lower) particle concentration. This can be done in two ways. For a given particle located at ${\bf{x}_0}$, the volume of its Vorono\"i cell $\altmathcal{V}$ can be multiplied by the concentration at the particle position $C({\bf{x}_0})$, i.e., $\altmathcal{V} C(\bf{x}_0)$, or by the average concentration in the entire cell that contains that particle $\langle C({\bf x}) \rangle_\altmathcal{V}$, i.e., $\altmathcal{V} \langle C \rangle_\altmathcal{V}$. We thus computed PDFs using these two compensated cell volumes, which are shown in Fig.~\ref{fig:cmap_VK}(c) along with the PDF of $\altmathcal{V}$ without any compensation. The overall shape of the PDFs remains similar, with the heavy tails that deviate from the RPP being preserved. The second compensation, using the mean concentration in each cell, does not change significantly the distribution of large volumes, but does decrease slightly the probability of finding small volumes. Even so, the compensated PDFs indicate that large-scale sampling effects as reported in \cite{Machicoane_2014} play a smaller role in our case, and confirm that the clustering observed in the experiment is at least partially associated to inertial effects. As a reference, the standard deviation of the Vorono\"i volumes compensated by $\langle C \rangle_\altmathcal{V}$ is $\sigma_{\altmathcal{V}_{\langle C \rangle}}/\sigma_{\text{RPP}} \approx 2.73$, which is slightly larger than $\sigma_\altmathcal{V}/\sigma_{\text{RPP}} \approx 2.33$ for the uncompensated Vorono\"i volumes in the experiment (see table \ref{tab:parameters}), and closer to the reference value for the minimal model of inertial particles in the TG DNSs.
\\

\subsection{Cluster volumes probability distribution}

\begin{figure}
    {\includegraphics[width=\textwidth ]{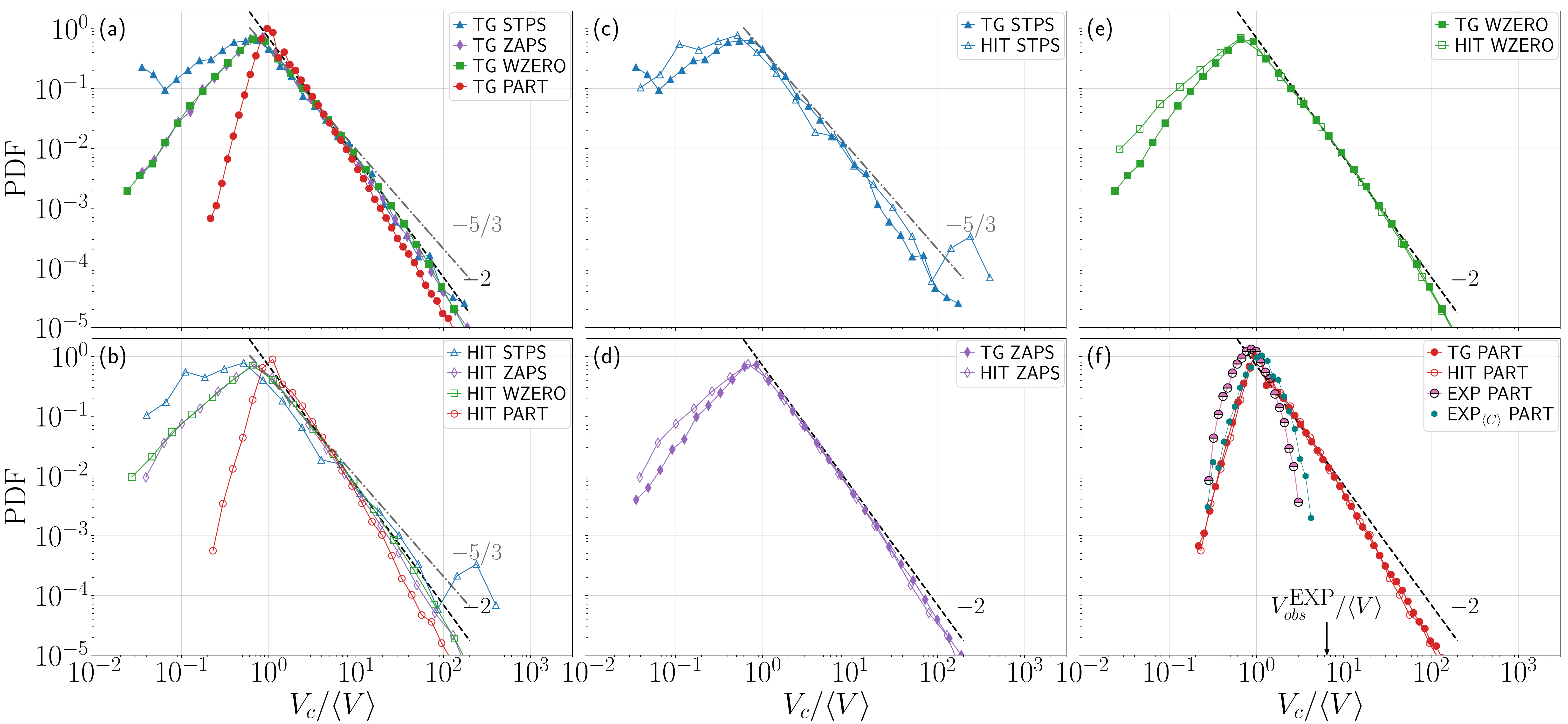}}
	\caption{PDFs of the cluster volumes normalized by the average Vorono\"{i} volume, $V_c/\langle V \rangle$, for STPS, ZAPS, WZERO and PART, in DNSs and EXP. Labels for the panels are as in Fig.~ \ref{fig:vor_volumes}. Panel (f) also shows the EXP data when corrected by the mean concentration per volume $\langle C \rangle$. The normalized observation volume in EXP is indicated by an arrow in the same panel. Power laws with exponents $-5/3$ and $-2$ are shown as references.}
  \label{fig:clu_volumes}
\end{figure}

Figure \ref{fig:clu_volumes} shows the PDFs of the volumes of the clusters $V_c$ (formed by at least two adjacent cells with $\altmathcal{V}<\altmathcal{V}_c$, where $\altmathcal{V}_c$ is the first crossing between the PDF of $\altmathcal{V}$ with the RPP, see \citet{Monchaux_2010}), for all vector fields nulls and for the particles, in the DNSs and EXP. From \citet{Mora_2021} we expect a $V_c^{-5/3}$ scaling for STPS, and $V_c^{-2}$ for ZAPS and WZEROs, resulting from the fractal nature of the spatial distribution of these zeros. Hence, the presence or not of these power laws provides information on the clusters geometry and their fractality (see \citet{Obligado_2014} and \citet{Mora_2021} for simple models and a review of the origin of these power laws). For all nulls, HIT shows a slight excess of smaller volume clusters when compared with TG, but overall the PDFs and scalings are similar. 

Figure \ref{fig:clu_volumes}(f) also shows the PDFs of $V_c$ for particles in the DNSs and EXP (in the latter case, with and without corrections by the mean average concentration per volume), which are expected to follow a $-2$ power-law scaling, with a shorter range as a result of the large $\textrm{St}$ considered. Clusters of particles in the experiment have different PDFs than in TG or HIT. The right tails of the PDFs in the experiment, associated with volumes $V_c > \langle V \rangle$, present a sharp drop as the cluster size cannot be larger than the observation volume ($V_{obs}^\text{EXP}/\langle V \rangle$ is indicated by the arrow in the figure). On the other hand, {in the PDF of uncompensated EXP volumes,} the probability of having small clusters (with $V_c < \langle V \rangle$) is larger than in the DNSs. We verified that this is not an effect of sub-sampling or boundaries, so this difference underlines limitations in the scope of the toy model considered in Eq.~\eqref{eq:iner} and, ultimately, in the effect of the mean large-scale flow as captured by the model, either from missing physical effects or from preferential sampling. Neglected interactions between particles is not the probable cause, as the volumetric loading ratio is $\lesssim 2\times 10^{-4}$. However, particles in the DNSs, while subjected to Stokes drag, are point particles that can only sense pressure gradients indirectly through the carrier velocity field. On the other hand, finite-size particles are expected to sense the pressure gradient directly as a force acting on their surface, as well as velocity gradients (and thus, shear stresses) that the point particles in the DNSs cannot sense. Pressure gradients have already been pointed out as important in the understanding of the formation of clusters of finite-size particles \citep{Fiabane_2012}. Finally, when compensating each EXP particle volume by the mean concentration, the probability of finding smaller clusters decreases and becomes closer to that of the model, suggesting that the small preferential concentration induced by the large-scale flow geometry plays a role here. This is confirmed by the fact that small and large clusters in TG and EXP flows occur at slightly different places, which could also explain why their PDFs look differently. 

In table \ref{tab:parameters} a comparison of the mean linear cluster size $\langle V_c\rangle^{1/3}$ for all quantities is given, normalized by $\eta$. The mean linear cluster sizes of WZERO and ZAPS are similar for TG and HIT, but STPS clusters are slightly larger for HIT, a difference that may be related to the fact that the HIT forcing scale is larger than in TG, and thus larger clusters may appear. The mean linear cluster size of particles in EXP is much larger than in TG or HIT. Even when conditioning the data as discussed in Sec.~\ref{sec:biases} (that is, sub-sampling the number of particles in the DNS, and considering boundary effects), or as in Sec.~\ref{sec:pref} (i.e., correcting by the particles concentration), we observe that clusters in the experiment are still 3 times larger, on average, than in TG. This difference is associated to the finite volume of the particles in EXP. Note that finite-size particles can only form clusters with a minimum linear size of order $4R$, which for EXP data is greater than $120\eta$. In other words, some clusters in TG cannot be realized in the experiment without the actual particles overlapping.

\begin{figure}
    {\includegraphics[width=\textwidth]{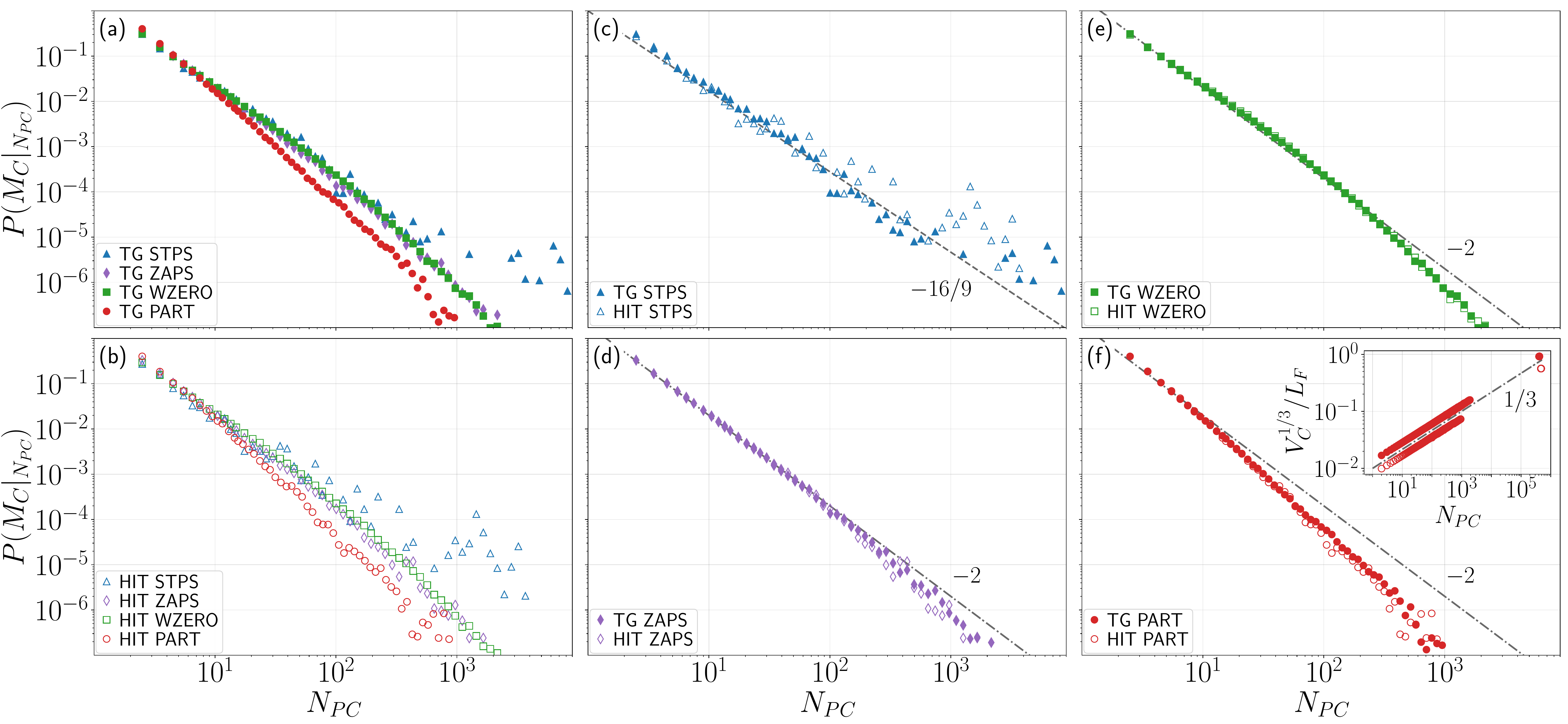}}
	\caption{Probability conditioned to having clusters with $N_{PC}$ points for STPS, ZAPS, WZERO, and PART in DNSs. Labels for all panels are as in Fig.~\ref{fig:vor_volumes}. The linear cluster size, normalized by $L_F$ and as a function of $N_{PC}$ is shown in the inset of panel (f). Several power laws are shown as references.}
  \label{fig:histogram_nc_1}
\end{figure}

The PDFs of the number of clusters $M_c$ with $N_{PC}$ points inside each cluster are shown for the DNSs in Fig.~\ref{fig:histogram_nc_1}. These PDFs follow a power-law behavior with an exponent close to $-16/9$ for STPS, and $-2$ for ZAPS and WZERO, as expected from \citet{Mora_2021}. Figure \ref{fig:histogram_nc_1}(c) indicates that the probability of having clusters of STPS with a larger number of nulls is larger for HIT than for TG, consistently with the larger mean linear size of clusters of STPS in HIT. Inertial particles also follow a power law with an exponent close to $-2$ for clusters with $N_{PC} \lesssim 10$, a behavior consistent with the one reported by \citet{Yoshimoto_2007}. Overall, the observed scaling is similar for ZAPS, WZERO, and inertial particles, which is compatible with the sweep-stick mechanism \citep{Goto_2008, Coleman2009AUS_2009, PhysRevFluids.6.044605_2021}. The deviation from the power law observed for particles in Fig.~\ref{fig:histogram_nc_1}(f) might relate to the fact that particles with large Stokes number filter out small flow scales. Indeed, these kinds of particles are expected to be less responsive to small scale motions in the fluid, so clusters with larger number of particles around very dense nulls in the carrier flow are less likely to form. Finally, the inset in Fig.~\ref{fig:histogram_nc_1}(f) shows the linear cluster size normalized by $L_F$, as a function of the number of particles per cluster for the inertial particles. Note the cluster size grows with the number of particles with a $1/3$ scaling.

\section{Discussion}

Comparison of clustering in numerical simulations of homogeneous isotropic turbulence and Taylor-Green turbulence with von K\'arm\'an experiments provides a powerful tool to characterize geometrical and topological properties of turbulence, and of multi-phase flows. Moreover, the similarities between the Taylor-Green flow and the laboratory von K\'arm\'an flow allow for detailed studies of possible biases when considering limitations in experiments such as low number of particles, boundary effects, or the effect of preferential sampling of the large-scale flow by the particles.

Our results also show that 3D Vorono\"{i} tessellation constitutes a powerful tool to study the topology of nulls and clusters, shedding light on dominant effects in the dynamics of the flow. For the case of the Taylor-Green and von K\'arm\'an flows, these are: (1) A lower probability of finding very densely packaged zeros of the velocity field, the vorticity, and of the Lagrangian acceleration in the TG flow when compared with HIT. (2) A deviation in the scaling of the number of Vorono\"i volumes of STPS with $\altmathcal{V}$ in the TG flow, probably associated with the effect of the mean flow on the velocity auto-correlation function. (3) Similar clustering properties of a minimal model of inertial particles in TG and of neutrally buoyant Taylor-scale particles in von K\'arm\'an flows as in inertial particles in HIT, probably resulting from the negligible differences (for intermediate cell volumes $\altmathcal{V}$) in the statistics of ZAPS and WZERO observed in TG and HIT.

In the experiment, we find that particles with a size comparable to the Taylor microscale form clusters that are very similar in intensity and size distribution as the point particles in the simulations. This contrasts with previous results, for particles whose sizes were a fraction of $\lambda$, and that did not exhibit clustering \citep{Qureshi_2007, Fiabane_2012, Fiabane_2013}, while close to point particles in experiments were found to form clusters \citep{obligado2019}, as well as much larger particles \citep{Machicoane_2014, Machicoane_2016} were also found to cluster for different reasons. This points towards the possibility that intermediate, Taylor-scale neutrally buoyant particles, cluster as a result of inertial effects. 

These results are promising as they further confirm Taylor-Green and von K\'arm\'an flows share many geometrical and statistical properties, even when considering preferential concentration of particles in multi-phase turbulence. Moreover, they indicate that many clustering properties of HIT can be extended to multi-scale flows with a large-scale circulation, at least in cases in which turbulent fluctuations are large (as in the TG and VK flows). Further studies will consider other forcing mechanisms and different large-scale geometries.

\begin{acknowledgments}
This work has been partially supported by ECOS-Sud Project No.~A18ST04. S.A., F.Z., P.J.C., and P.D.M.~acknowledge support from UBACYT Grant No.~20020170100508BA and PICT Grant No.~2018-4298. A.F. and M.O. acknowledge the LabEx Tec21 (Investissements d'Avenir - Grant Agreement $\#$ ANR-11-LABX-0030). The authors thank Daniel Mora for help with data analysis of nulls calculation. The authors gratefully acknowledge suggestions by an anonymous referee of studying the spatial concentration map of the particles in the experiment. The authors report no conflict of interest.
\end{acknowledgments}

\appendix
\section{Models with other forces and their effect on clustering}
\label{appendix}

\begin{figure}
    \begin{center}
    {\includegraphics[width=0.6\textwidth]{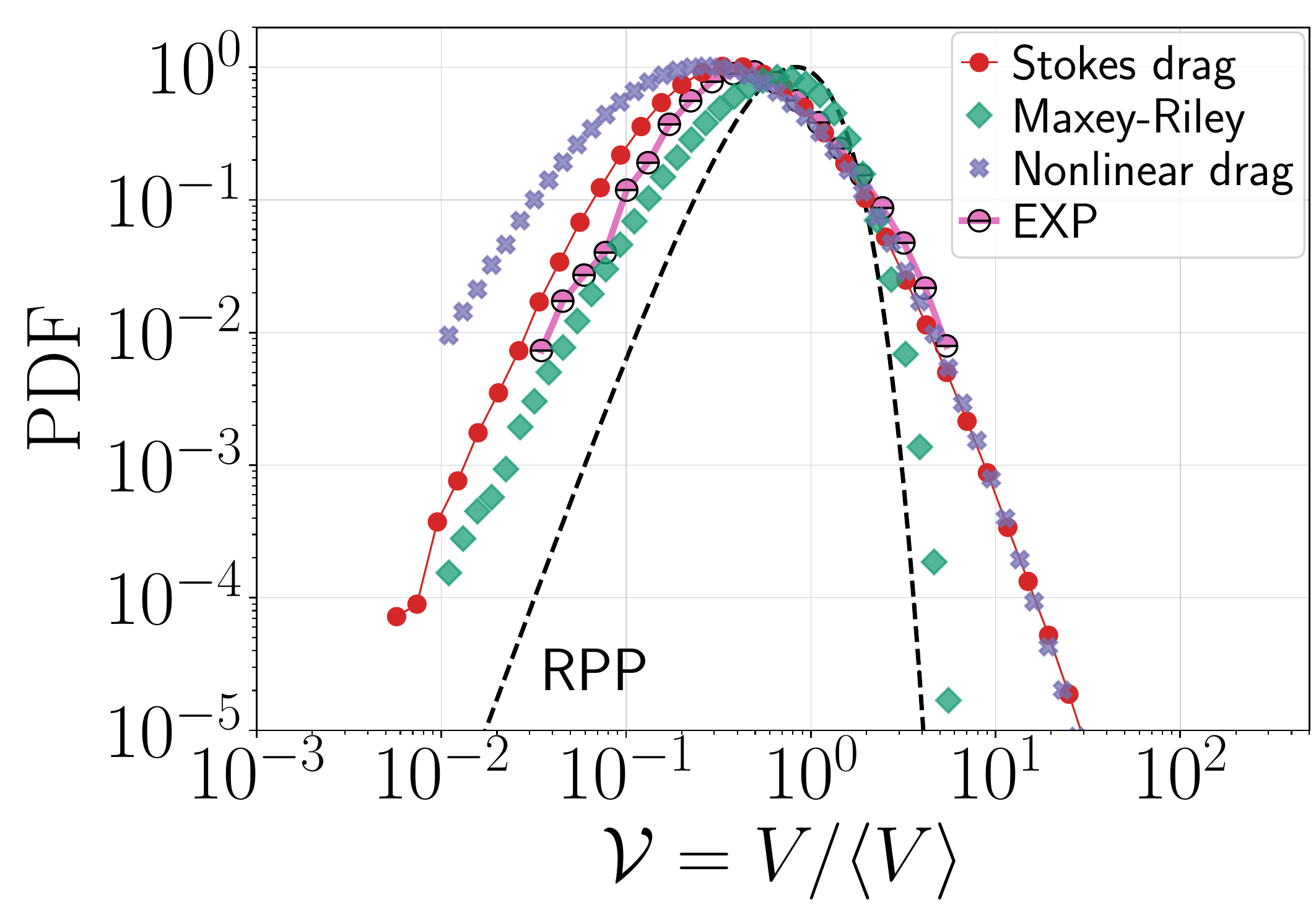}}
	\end{center}
	\caption{PDFs of normalized Vorono\"i volumes for inertial particles in a TG flow considering only Stokes drag as in Eq.~\eqref{eq:iner}, the Maxey-Riley equation (with Stokes drag, gravity, particle acceleration and added mass effects), and nonlinear drag. EXP data is shown as well for comparison.}
  \label{fig:ipart_models}
\end{figure}

In this appendix we briefly consider other possible models for the equations of motion of particles in the simulations. On the one hand we consider particles subjected to a nonlinear drag force as in \citet{Wang_1993}, whose equations of motion read
\begin{equation}
    \dot{\bf x}_p = {\bf v}(t) , \qquad
    \dot{\bf v} = \frac{1 + 0.15~\textrm{Re}_p^{0.687}}{\tau_p}~ (\mathbf{u} - \mathbf{v}), \label{eq:nonlindrag}
\end{equation}
where $\textrm{Re}_p = \sqrt{18\tau_p/\nu}~|\mathbf{u} - \mathbf{v}|$. On the other hand we consider the Maxey-Riley equation \citep{Maxey_1983} up to terms linear in the particle's radius, that is, we consider the Stokes drag, gravity, and particle acceleration and added mass effects (and therefore neglecting Fax\'en corrections and the Basset-Boussinesq history term). Thus, particles evolve according to
\begin{equation}
    \dot{\bf x}_p = {\bf v}(t) , \qquad
    \dot{\bf v} = \frac{1}{\tau_p} ({\bf u} - {\bf v})~ -~g \frac{1 - \gamma}{1 + \gamma/2} \hat{z} ~ + ~ \frac{3}{2} \frac{\gamma}{1 + \gamma/2}~ \frac{D\mathbf{u}}{Dt} ,
    \label{eq:MR}
\end{equation}
where $g$ is the gravitational acceleration, $D/Dt = \partial/\partial t + {\bf u} \cdot \boldsymbol{\nabla}$, and $\gamma = \rho_f/\rho_p$. Particles whose dynamics are governed by either Eq.~\eqref{eq:nonlindrag} or \eqref{eq:MR} were evolved in the turbulent flow with Taylor-Green mechanical forcing with a spatial resolution of $768^3$ grid points, following the same procedures as previously described for the TG DNSs. A total of $10^6$ particles were integrated. For the sake of comparison, and as both equations are missing other forces \citep[see, e.g.,][]{Calzavarini_2008, Calzavarini_2009}, we use in all cases the effective particle response time $\tau_p$ instead of the viscous relaxation time $\tau_p^v$, and in Eq.~\eqref{eq:MR} we use $\gamma = 1/1.02 \approx 0.98$ and the value of $g$ to match the experiment. We performed a Vorono\"i tessellation of the particles' positions, and the resulting cell volumes PDFs are shown in Fig.~\ref{fig:ipart_models}, along with the PDFs corresponding to particles evolving solely under Stokes drag as in Eq.~\eqref{eq:iner}, and the experimental data. As expected, considering particle acceleration and added mass effects in the Maxey-Riley equation results in less clustering: Eq.~\eqref{eq:MR} is well suited for smaller particles with small particle Reynolds number, which tend to cluster less \citep{Fiabane_2012, Fiabane_2013}. Note in particular how the statistics of voids (i.e., of large Vorono\"i cells) approaches the statistics of the RPP; the standard deviation of this PDF is $\sigma_\altmathcal{V} = 1.31 \sigma_\text{RPP}$. Considering nonlinear drag as in Eq.~\eqref{eq:nonlindrag} may be more relevant for particles of our size and with larger particle Reynolds number. This model results instead in more clusterization, with a PDF with  $\sigma_\altmathcal{V} = 2.92 \sigma_\text{RPP}$. The stronger clustering may be related to our choice of using an effective $\tau_p$. In the future, it may be of interest to compare nonlinear drag using the viscous response time of the particles (or other choices for the effective response time) against the simplified point particle model with the effective parameter in Eq.~\eqref{eq:iner}. Note in particular that the results in Fig.~\ref{fig:ipart_models} do not necessarily imply that Eq.~\eqref{eq:iner} is a better model for Taylor-scale neutrally buoyant particles, as to that end a more detailed comparison would be needed.

\bibliographystyle{jfm}
\bibliography{ms}

\end{document}